\documentclass[11pt]{article}
\usepackage{amsmath,amssymb}
\usepackage{fullpage}
\usepackage{natbib}
\usepackage{epsfig}
\usepackage{mathptm}
\usepackage{subfigure}

\newcommand{\ps}{\Psi}
\newcommand{\pst}{\tilde{\Psi}}
\newcommand{\pt}{\tilde{P}}
\newcommand{\yt}{\tilde{Y}}

\newcommand{\mat}{\left( \begin{array}{cc}}
\newcommand{\rix}{\end{array} \right)}
\newcommand{\vk}{{\vec{k}}}
\newcommand{\vx}{{\vec{x}}}
\newcommand{\vy}{{\vec{y}}}
\newcommand{\Z}{\mathbb{Z}}
\newcommand{\C}{\mathbb{C}}

\newcommand{\ord}{{\cal O}}
\newcommand{\eps}{\epsilon}
\newcommand{\p}{\partial}
\newcommand{\dt}{{\rm d}t}
\newcommand{\dx}{{\rm d}x}
\newcommand{\dy}{{\rm d}y}
\newcommand{\domega}{{\rm d}\omega}
\newcommand{\dk}{{\rm d}k}
\newcommand{\dl}{{\rm d}l}
\newcommand{\dbeta}{{\rm d}\beta}
\newcommand{\e}{{\rm e}}
\newcommand{\eqdef}{\stackrel{\text{eq}}{=}}
\newcommand{\norm}[1]{\left\| #1 \right\|}
\newcommand{\id}{{\bf 1}}
\newcommand{\oP}{\bar{P}}
\newcommand{\abs}[1]{\left| #1 \right|}
\newcommand{\sgn}{{\rm sgn}}

\newtheorem{definition}{Definition}
\newtheorem{theorem}{Theorem}

\begin{document}

\title{Quantum Walks on the Hypercube}

\author{\begin{minipage}[t]{3in}
    \begin{center}
      \textsc{Cristopher Moore} \\
      \begin{small}
        \begin{minipage}[t]{3in}
          \begin{center}
            Computer Science Department \\
            University of New Mexico, Albuquerque \\
            and the Santa Fe Institute, Santa Fe, New Mexico \\
            {\tt moore@cs.unm.edu}
          \end{center}
        \end{minipage}
      \end{small}
    \end{center}
  \end{minipage}
  \and \begin{minipage}[t]{3in}
    \begin{center}
      \textsc{Alexander Russell} \\
      \begin{small}
        \begin{minipage}[t]{3in}
          \begin{center}
            Department of Computer Science and Engineering\\
            University of Connecticut \\
            Storrs, Connecticut \\ 
            \texttt{acr@cse.uconn.edu}
          \end{center}
        \end{minipage}
      \end{small}
    \end{center}
  \end{minipage}}

  \maketitle

\begin{abstract}  
  Recently, it has been shown that one-dimensional quantum walks can
  mix more quickly than classical random walks, suggesting that
  quantum Monte Carlo algorithms can outperform their classical
  counterparts.  We study two quantum walks on the $n$-dimensional
  hypercube, one in discrete time and one in continuous time. In both
  cases we show that the quantum walk mixes in $(\pi/4)n$ steps,
  faster than the $\Theta(n \log n)$ steps required by the classical
  walk.  In the continuous-time case, the probability distribution is
  {\em exactly} uniform at this time.  More importantly, these walks
  expose several subtleties in the definition of mixing time for
  quantum walks.  Even though the continuous-time walk has an
  $\ord(n)$ instantaneous mixing time at which it is precisely uniform,
  it never approaches the uniform distribution when the stopping time
  is chosen randomly as in \cite{stoc::AharonovAKV2001}.  Our analysis
  treats interference between terms of different phase more carefully
  than is necessary for the walk on the cycle; previous general bounds
  predict an exponential, rather than linear, mixing time for the
  hypercube.
\end{abstract}

\vspace{-1ex}
\section{Introduction}

Random walks form one of the cornerstones of theoretical computer
science.  As algorithmic tools, they have been applied to a variety of
central problems, such as estimation of the volume of a convex body
\cite{jacm::DyerFK91,stoc::LovaszK1999}, approximation of the
permanent \cite{sicomp::JerrumS89,eccc::JerrumSV2000}, and discovery
of satisfying assignments for Boolean formulae
\cite{focs::Schoning1999}. Furthermore, the basic technical phenomena
appearing in the study of random walks (e.g., spectral decomposition,
couplings, and Fourier analysis) also support several other important
areas such as pseudorandomness and derandomization (see, e.g.,
\cite[({\S}9,{\S}15)]{AlonS:Probabilistic}).

The development of efficient \emph{quantum} algorithms for problems
believed to be intractable for (classical) randomized computation,
like integer factoring and discrete logarithm \cite{sicomp::Shor1997},
has prompted the investigation of \emph{quantum walks}. This is a
natural generalization of the traditional notion discussed above
where, roughly, the process evolves in a unitary rather than
stochastic fashion.

The notion of ``mixing time,'' the first time when the distribution
induced by a random walk is sufficiently close to the stationary
distribution, plays a central role in the theory of classical random
walks.  For a given graph, then, it is natural to ask if a quantum
walk can mix more quickly than its classical counterpart.  (Since a
unitary process cannot be mixing, we define a stochastic process from
a quantum one by performing a measurement at a given time or a
distribution of times.)  Several recent articles
\cite{stoc::AharonovAKV2001,stoc::AmbainisBNVW2001,quant-ph::NayakV2000}
have answered this question in the affirmative, showing, for example,
that a quantum walk on the $n$-cycle mixes in time $\ord(n \log n)$, a
substantial improvement over the classical random walk which requires
$\Theta(n^2)$ steps to mix.  Quantum walks were also defined
in~\cite{jcss::Watrous2001}, and used to show that undirected graph
connectivity is contained in a version of quantum LOGSPACE.  These
articles raise the exciting possibility that quantum Monte Carlo
algorithms could form a new family of quantum algorithms that work
more quickly than their classical counterparts.

Two distinct notions of quantum walks exist in the literature. The
first, introduced by
\cite{stoc::AharonovAKV2001,stoc::AmbainisBNVW2001,quant-ph::NayakV2000},
studies the behavior of a ``directed particle'' on the graph; we refer
to these as \emph{discrete-time} quantum walks. The second, introduced
by \cite{quant-ph::ChildsFG2001}, defines the dynamics by treating the
adjacency matrix of the graph as a Hamiltonian; we refer to these as
\emph{continuous-time} quantum walks. The landscape is further
complicated by the existence of two distinct notions of mixing time.
The first ``instantaneous'' notion
\cite{stoc::AmbainisBNVW2001,quant-ph::NayakV2000} focuses on
particular times at which measurement induces a desired distribution;
the second ``average'' notion \cite{stoc::AharonovAKV2001}, another
natural way to convert a quantum process into a stochastic one,
focuses on measurement times selected at random.

In this article, we analyze both the continuous-time and a
discrete-time quantum walk on the hypercube. In both cases, the walk
is shown to have an instantaneous mixing time at $(\pi/4) n$. Recall
that the classical walk on the hypercube mixes in time $\Theta(n \log
n)$, so that the quantum walk is faster by a logarithmic factor.
Moreover, in the discrete-time case the walk mixes in time less than
the diameter of the graph, since $\pi/4 < 1$; and, astonishingly, in
the continuous-time case the probability distribution at $t= (\pi/4)
n$ is \emph{exactly} uniform.  Both of these things happen due to a
marvelous conspiracy of destructive interference between terms of
different phase.

These walks show \textsl{i.)} a similarity between the two notions of
quantum walks, and \textsl{ii.)} a disparity between the two notions
of quantum mixing times.  As mentioned above, both walks have an
instantaneous mixing time at time $(\pi/4)n$.  On the other hand, we
show that there is \emph{no} time at which the continuous walk
approaches the uniform distribution in the sense
of~\cite{stoc::AharonovAKV2001}.  Thus there are some real subtleties
involved in defining mixing times for quantum walks.

The analysis of the hypercubic quantum walk exhibits a number of
features markedly different from those appearing in previously studied
walks.  In particular, the dimension of the relevant Hilbert space is,
for the hypercube, exponential in the length of the desired walk,
while in the cycle these quantities are roughly equal.  This requires
that interference be handled in a more delicate way than is required
for the walk on the cycle; in particular, the general bound of
\cite{stoc::AharonovAKV2001} predicts an exponentially large mixing
time for the discrete-time walk.

We begin by defining quantum walks and discussing various notions of
mixing time.  We then analyze the two quantum walks on the hypercube
in Sections~\ref{sec:discrete-walk}
and~\ref{sec:continuous-walk}. (Most of the technical details for the
discrete-time walk are relegated to an appendix.)  Finally, in
Section~\ref{sec:mixingperiods}, we discuss mixing times in the sense
of \cite{stoc::AharonovAKV2001}.

\subsection{Quantum walks and mixing times}

Any graph $G = (V,E)$ gives rise to a familiar Markov chain by
assigning probability $1/d$ to all edges leaving each vertex $v$ of
degree $d$. Let $P_{u}^t(v)$ be the probability of visiting a vertex
$v$ at step $t$ of the random walk on $G$ starting at $u$.  If $G$ is
undirected, connected, and not bipartite, then $\lim_{t \to \infty} P_u^t$
exists\footnote{In fact, this limit exists under more general
  circumstances; see e.g.\ \cite{MotwaniR:Randomized}.} and is
independent of $u$. A variety of well-developed techniques exist
for establishing bounds on the rate at which $P_u^t$ achieves this
limit (e.g.,\ \cite{Vazirani:Rapidly}); if $G$ happens to be the
Cayley graph of a group (as are, for example, the cycle and the
hypercube), then techniques from Fourier analysis can be applied (see
\cite{Diaconis:Group}).  Below we will use some aspects of this
approach, especially the Diaconis-Shahshahani bound on the total
variation distance~\cite{DiaconisS:Generating}.

For simplicity, we restrict our discussion to quantum walks on Cayley
graphs; more general treatments of quantum walks appear in
\cite{stoc::AharonovAKV2001,quant-ph::ChildsFG2001}. Before describing
the quantum walk models we set down some notation.

\noindent
\textbf{Notation.} For a group $G$ and a set of generators $\Gamma$ such
that $\Gamma = \Gamma^{-1}$, we let $X(G,\Gamma)$ denote the undirected Cayley
graph of $G$ with respect to $\Gamma$. For a finite set $S$, we let $L(S)
= \{ f : S \to \C \}$ denote the collection of $\C$-valued functions on
$S$. This is a Hilbert space under the natural inner product $\langle f | g
\rangle = \sum_{s \in S} f(s) \,g(s)^*$. For a Hilbert space $V$, an operator
$U: V \to V$ is \emph{unitary} if for all $\vec{v},\vec{w} \in V$, $\langle
\vec{v} | \vec{w} \rangle = \langle U \vec{v} | U \vec{w}\rangle$; if $U$ is
represented as a matrix, this is equivalent to the condition that
$U^\dagger = U^{-1}$ where $\dagger$ denotes the Hermitian conjugate.

There are two natural quantum walks that one can define for such
graphs, which we now describe.
\begin{description}
\item{\textbf{The discrete-time walk.}}  This model, introduced by
  \cite{stoc::AharonovAKV2001,stoc::AmbainisBNVW2001,quant-ph::NayakV2000},
  augments the space $L(G)$ with a {\em direction space}, each basis
  vector of which corresponds one of the generators in $\Gamma$. A step
  of the walk then consists of the composition of two unitary
  transformations; a {\em shift} operator which leaves the direction
  unchanged while moving the particle in the appropriate direction, and
  a {\em local transformation} which operates on the direction while
  leaving the position unchanged.  To be precise, the quantum walk on
  $X(G,\Gamma)$ is defined on the space $L(G \times \Gamma) \cong L(G)
  \otimes L(\Gamma)$.  Let $\{ \delta_\gamma \mid \gamma \in \Gamma \}$
  be the natural basis for $L(\Gamma)$, and $\{ \delta_g | \,g \in G \}$ the
  natural basis for $L(G)$.  Then the shift operator is $S: (\delta_g
  \otimes \delta_\gamma) \mapsto (\delta_{g\gamma} \otimes
  \delta_\gamma)$, and the local transformation is $\check{D} = \id
  \otimes D$ where $D$ is defined on $L(\Gamma)$ alone and $\id$ is the
  identity on $L(G)$.  Then one ``step'' of the walk corresponds to the
  operator $U = \check{D}V$.  If we measure the position of the
  particle, but not its direction, at time $t$, we observe a vertex $v$
  with probability $P_t(v) = \sum_{\gamma \in \Gamma} \left|
    \left\langle U^t \psi_0 \mid \delta_v \otimes
      \delta_\gamma\right\rangle \right|^2$ where $\psi_0 \in L(G \times
  \Gamma)$ is the initial state.
\item{\textbf{The continuous-time walk.}}  This model, introduced by
  \cite{quant-ph::ChildsFG2001}, works directly with $L(G)$, the
  Hilbert space of $\C$-valued functions on $G$: $L(G) = \{ f : G \to
  \C\}$.  The walk evolves by treating the adjacency matrix of the
  graph as a Hamiltonian and using the Schr\"{o}dinger equation.
  Specifically, if $H$ is the adjacency matrix of $X(G,\Gamma)$, the
  evolution of the system at time $t$ is given by $U_t$, where $U_t
  \eqdef \e^{iHt}$ (here we use the matrix exponential, and $U_t$ is
  unitary since $H$ is real and symmetric).  Then if we measure the
  position of the particle at time $t$, we observe a vertex $v$ with
  probability $P_t(v) = \left| \left\langle U_t \psi_0 | e_v \right\rangle
  \right|^2$ where $\psi_0$ is the initial state.
\end{description}

In both cases we start with an initial wave function concentrated at
a single vertex $u$. For the continuous-time walk, this
corresponds to a wave function
\[
\psi_u(v) = \langle \psi_u | \delta_v \rangle
= \begin{cases} 1 & \textrm{if}\; u = v, \\
  0 & \textrm{otherwise.}\end{cases}
\]
For the discrete-time walk, we start with a uniform superposition over
all possible directions,
\[
\psi_u(v,\gamma) 
= \left\langle \psi_u | \,e_v \otimes e_\gamma \right\rangle 
= \begin{cases} 1/\sqrt{|\Gamma|} & \textrm{if}\; u = v, \\
  0 & \textrm{otherwise.}\end{cases}
\]

In order to define a discrete quantum walk, one must select a local
operator $D$ on the direction space.  In principle, this introduces
some arbitrariness into the definition.  However, if we wish $D$ to
respect the permutation symmetry of the $n$-cube, and if we wish to
maximize the operator distance between $D$ and the identity, we show
in Appendix~\ref{app:grover} that we are forced to choose Grover's
diffusion operator \cite{stoc::Grover96}, which we recall below. We
call the resulting walk the ``symmetric discrete-time quantum walk''
on the $n$-cube.  (Watrous \cite{jcss::Watrous2001} also used Grover's
operator to define quantum walks on undirected graphs.)

(Since for large $n$ Grover's operator is close to the identity
matrix, one might imagine that it would take $\Omega(n^{1/2})$ steps
to even change direction, giving the quantum walk a mixing time of
$\approx n^{3/2}$, slower than the classical random walk.  However,
like many intuitions about quantum mechanics, this is simply wrong.)

Since the evolution of the quantum walk is governed by a unitary
operator rather than a stochastic one, unless $P_t$ is constant for
all $t$, there can be no ``stationary distribution'' $\lim_{t \to
\infty} P_t$.  In particular, for any $\epsilon > 0$, there are
infinitely many (positive, integer) times $t$ for which $\norm{U^t -
\id} \leq \epsilon$ so that $\norm{ U^t \psi_u - \psi_u} \leq
\epsilon$ and $P_t$ is close to the initial distribution.  However,
there may be particular stopping times $t$ which induce distributions
close to, say, the uniform distribution, and we call these
\emph{instantaneous mixing times}:

\begin{definition}
  We say that $t$ is an \emph{$\eps$-instantaneous mixing time}
  for a quantum walk if $\norm{P_t - U} \leq \epsilon$, where
\[ \norm{A - B} = \frac{1}{2} \sum_v |A(v) - B(v)| \]
denotes total variation distance and $U$ denotes the uniform
distribution.
\end{definition}

For these walks we show:
\begin{theorem}\label{thm:discrete}
  For the symmetric discrete-time quantum walk on the $n$-cube, $t =
  \lceil k (\pi/4) n \rceil$ is an $\epsilon$-instantaneous mixing
  time with $\epsilon = \ord(n^{-7/6})$ for all odd $k$.
\end{theorem}
and, even more surprisingly,
\begin{theorem}\label{thm:continuous}
  For the continuous-time quantum walk on the $n$-hypercube, $t = k
  (\pi/4) n$ is a $0$-instantaneous mixing time for all odd $k$.
\end{theorem}
Thus in both cases the mixing time is $\Theta(n)$, as opposed to
$\Theta(n \log n)$ as it is in the classical case.

Aharonov et al.\ \cite{stoc::AharonovAKV2001} define another natural
notion of mixing time for quantum walks, in which the stopping time
$t$ is selected uniformly from the set $\{0,\ldots,T-1\}$.  They show
that the distributions $\oP_T = \frac{1}{T} \sum_{t=0}^{T-1} P_t$
\emph{do} converge as $T \to \infty$ and study the rate at which this
occurs.  For a continuous random walk, we analogously define the
distribution $\oP_T(v) = (1/T) \int_{0,T} P_t(v) \,\textrm{d}t$.  Then
we call a time at which the resulting distribution $\oP_T$ is close to
uniform an \emph{average mixing time}:

\begin{definition}
  We say that $T$ is an \emph{$\epsilon$-average mixing time}
  for a quantum walk if $\norm{\oP_T - U} \leq \epsilon$.
\end{definition}

The exact relationship between instantaneous and average mixing times
is unclear.  In fact, while the continuous walk on the hypercube
possesses $0$-instantaneous mixing times at all odd multiples of
$(\pi/4)n$, the limiting distribution of $\oP_T$ is \emph{not} the
uniform distribution, and we will show that an $\epsilon > 0$ exists such
that {\em no} time is an $\epsilon$-average mixing time.  For the
discrete-time walk, the limiting distribution \emph{is} uniform and we
show that the general bound given in~\cite{stoc::AharonovAKV2001}
predicts an exponential, rather than linear, average mixing time for
the hypercube.

\section{The symmetric discrete-time walk}
\label{sec:discrete-walk}

In this section we prove Theorem~\ref{thm:discrete}.  We treat the
$n$-cube as the Cayley graph of $\Z_2^n$ with the regular basis
vectors $\vec{e}_i = (0, \ldots, 1, \ldots, 0)$ with the $1$ appearing in the
$i$th place.  Then the discrete-time walk takes place in the Hilbert
space $L(\Z_2^n \times [n])$ where $[n] = \{1, \ldots, n\}$.  Here the first
component represents the position of the particle in the hypercube,
and the second component represents the ``direction'' currently
associated with the particle.

As in \cite{stoc::AharonovAKV2001,quant-ph::NayakV2000}, we will not
impose a group structure on the direction space, and will Fourier
transform only over the position space.  For this reason, we will
express an element $\psi$ in $L(\Z_2^n) \otimes L([n])$ as a function
$\Psi: \Z_2^n \to \C^n$, where the $i$th coordinate of $\Psi(\vec{x})$
is the projection of $\psi$ into $\delta_{\vec{x}} \otimes \delta_i$,
i.e.\ the complex amplitude of the particle being at position
$\vec{x}$ with direction $i$.  The Fourier transform of such an
element $\Psi$ is $\pst: \Z_2^n \to \C^n$, where
\[ \pst(\vk) = \sum_\vx (-1)^{\vk \cdot \vx} \,\Psi(\vec{x}). \]
Then the shift operator for the hypercube is
\[ S: \,\ps(x) \mapsto \sum_{i=1}^n \pi_i \ps(\vec{x} \oplus \vec{e}_i) \]
where $\vec{e}_i$ is the $i$th basis vector in the $n$-cube, and
$\pi_i$ is the projection operator for the $i$th direction.  The reason
for considering the Fourier transform above is that the shift operator
is locally diagonal in this basis: specifically it maps $\pst(\vk) \mapsto
S_\vk \,\pst(\vk)$ where
\begin{small}
\[ S_\vk = \left( \begin{array}{cccc}
(-1)^{k_1} & & & 0 \\
& (-1)^{k_2} & & \\
& & \ddots & \\
0 & & & (-1)^{k_n}
\end{array} \right) \]
\end{small}
For the local transformation, we use Grover's diffusion operator
on $n$ states, $D_{ij} = 2/n - \delta_{ij}$.

The advantage of Grover's operator is that, like the $n$-cube itself,
it is permutation symmetric.  We use this symmetry to rearrange
$U_\vk=S_\vk D$ to put the negated rows on the bottom,
\begin{small}
\[ U_\vk = \left( \begin{array}{ccc|ccc}
2/n-1  & 2/n    & \cdots &        &        & \\
2/n    & 2/n-1  &        &        &  2/n   & \\
\vdots &        & \ddots &        &        & \\ \hline
       &        &        & -2/n+1 & -2/n   & \cdots \\
       & -2/n   &        & -2/n   & -2/n+1 & \\
       &        &        & \vdots &        & \ddots
\end{array} \right)
\]
\end{small}
where the top and bottom blocks have $n-k$ and $k$ rows respectively;
here $k$ is the Hamming weight of $\vk$.

The eigenvalues of $U_\vk$ then depend only on $k$.  Specifically,
$U_\vk$ has the eigenvalues $+1$ and $-1$ with multiplicity $k-1$ and
$n-k-1$ respectively, plus the eigenvalues $\lambda, \lambda^*$ where
\[ \lambda = 1 - \frac{2k}{n} + \frac{2i}{n} \sqrt{k(n-k)}
           = \e^{i\omega_k} \]
and $\omega_k \in [0,\pi]$ is described by 
\[ \cos \omega_k = 1-\frac{2k}{n}, \quad 
   \sin \omega_k = \frac{2}{n} \sqrt{k(n-k)} 
\]
Its eigenvectors with eigenvalue $+1$ span the $(k-1)$-dimensional
subspace consisting of vectors with support on the $k$ ``flipped''
directions that sum to zero, and similarly the eigenvectors with
eigenvalue $-1$ span the $(n-k-1)$-dimensional subspace of vectors on
the $n-k$ other directions that sum to zero.  We call these the
\emph{trivial} eigenvectors.  The eigenvectors of $\lambda,\lambda^* =
\e^{\pm i \omega_k}$ are
\[ v_k, v_k^* = \frac{1}{\sqrt{2}} 
\Bigl( \underbrace{\frac{\mp i}{\sqrt{n-k}}}_{n-k} ,
\underbrace{\frac{1}{\sqrt{k}}}_k \Bigr).
\]
We call these the \emph{non-trivial} eigenvectors for a given $\vk$.
Over the space of positions and directions these eigenvectors are
multiplied by the Fourier coefficient $(-1)^{\vk \cdot \vx}$, so as a
function of $\vx$ and direction $1 \leq j \leq n$ the two non-trivial
eigenstates of the entire system, for a given $\vk$, are
\[ v_\vk(\vx,j) = (-1)^{\vk \cdot \vx} \,\frac{2^{-n/2}}{\sqrt{2}} 
\times \left\{ \begin{array}{ll}
1/\sqrt{k} & \mbox{ if } \vk_j = 1 \\
-i/\sqrt{n-k} & \mbox{ if } \vk_j = 0 
\end{array} \right.
\]
with eigenvalue $\e^{i\omega_k}$, and its conjugate $v_\vk^*$ with
eigenvalue $\e^{-i\omega_k}$.

We take for our initial wave function a particle at the origin
$u=(0,\ldots,0)$ in an equal superposition of directions.  Since its
position is a $\delta$-function in real space it is uniform in Fourier
space as well as over the direction space, giving
\[ \pst_0(\vk) = \frac{2^{-n/2}}{\sqrt{n}} (1,\ldots,1) \]
This is perpendicular to all the trivial eigenvectors, so their
amplitudes are all zero.  The amplitude of its component along the
non-trivial eigenvector $v_\vk$ is
\begin{equation}
a_\vk = \langle \ps_0 | v_\vk \rangle = \frac{2^{-n/2}}{\sqrt{2}} 
\left( \sqrt{\frac{k}{n}} - i \sqrt{1-\frac{k}{n}} \right) 
\label{ai}
\end{equation}
and the amplitude of $v_\vk^*$ is $a_\vk^*$.  Note that 
$|a_\vk|^2 = 2^{-n}/2$, so a particle is equally likely to appear
in either non-trivial eigenstate with any given wave vector.  

At this point, we note that there are an exponential number of
eigenvectors in which the initial state has a non-zero amplitude.  In
Section~\ref{sec:mixingperiods}, we show that the general bound of
Aharonov et al.\ \cite{stoc::AharonovAKV2001} predicts an exponential
mixing time. In general, this bound performs poorly whenever the
number of important eigenvalues is greater than the mixing time.

Instead, we will use the Diaconis-Shahshahani bound on the total
variation distance in terms of the Fourier coefficients of the
probability~\cite{Diaconis:Group}.  If $P_t(\vx)$ is the probability
of the particle being observed at position $\vx$ at time $t$, and $U$
is the uniform distribution, then the total variation distance
is bounded by
\begin{equation}
  \norm{P_t-U}^2 \leq \frac{1}{4} \sum_{\scriptsize \begin{array}{c}
  \vk \neq (0,\ldots,0) \\
  \vk \neq (1,\ldots,1) \end{array}}
  \left| \pt_t(\vk) \right|^2
  = \frac{1}{4} \sum_{k=1}^{n-1} \binom{n}{k} \abs{\pt_t(k)}^2.
  \label{dcbound}
\end{equation}
Here we exclude both the constant term and the parity term
$\vk=(1,\ldots,1)$; since our walk changes position at every step,
we only visit vertices with odd or even parity at odd or even
times respectively.  Thus $U$ here means the uniform distribution
with probability $2^{n-1}$ on the vertices of appropriate parity.

To find $\pt_t(\vk)$, we first need $\pst_t(\vk)$.  As Nayak and
Vishwanath \cite{quant-ph::NayakV2000} did for the walk on the line,
we start by calculating the $t$th matrix power of $U_\vk$.  This is
\begin{small}
\[ 
U_\vk^t = \left( \begin{array}{ccc|ccc}
a+(-1)^t & a         & \cdots &          &          & \\
a        & a+(-1)^t  &        &          &     c    & \\
\vdots   &           & \ddots &          &          & \\ \hline
         &           &        & b-(-1)^t & b        & \cdots \\
         &    -c     &        & b        & b-(-1)^t & \\
         &           &        & \vdots   &          & \ddots
\end{array} \right)
\]
\end{small}
where 
\[ a = \frac{\cos \omega_k t - (-1)^t}{n-k}, \quad
   b = \frac{\cos \omega_k t + (-1)^t}{k}, \quad
\mbox{and} \quad
   c = \frac{\sin \omega_k t}{\sqrt{k(n-k)}} \]
Starting with the uniform initial state, the wave function after $t$ 
steps is
\begin{equation}
 \pst_t(\vk) = \frac{1}{\sqrt{n}} \Bigl( 
\underbrace{\cos \omega_k t + \sqrt{\frac{k}{n-k}} \sin \omega_k t}_{n-k} ,
\,
\underbrace{\cos \omega_k t - \sqrt{\frac{n-k}{k}} \sin \omega_k t}_k 
\Bigr)
\label{pst}
\end{equation}

We could, at this point, calculate $\ps_t(\vx)$ by Fourier transforming
this back to real space.  However, this calculation turns out to be
significantly more awkward than calculating the Fourier transform of
the probability distribution, $\pt_t(\vk)$, which we need to apply the
Diaconis-Shahshahani bound.  Since $P_t(\vx) = \ps_t(\vx) \ps_t(\vx)^*$,
and since multiplications in real space are convolutions in Fourier
space, we perform a convolution over $\Z_2^n$:
\[ \pt_t(\vk) = \sum_{\vk'} \pst_t(\vk') \cdot \pst_t(\vk \oplus \vk') \]
where the inner product is defined on the direction space, $u \cdot v
= \sum_{i=1}^n u_i v_i^*$.  We write this as a sum over $j$, the
number of bits of overlap between $\vk'$ and $\vk$, and $l$, the
number of bits of $\vk'$ outside the bits of $\vk$ (and so overlapping
with $\vk \oplus \vk'$).  Thus $\vk'$ has weight $j+l$, and $\vk
\oplus \vk'$ has weight $k-j+l$.

Calculating the dot product $\pst_t(\vk') \cdot \pst_t(\vk \oplus
\vk')$ explicitly from Equation~\ref{pst} as a function of these
weights and overlaps, we have
\begin{equation}
\pt_t(k) = \frac{1}{2^n} \sum_{j=0}^k \sum_{l=0}^{n-k}
\binom{k}{j} \binom{n-k}{l} 
\Biggl[ \,
\cos \omega_{j+l} t \,\cos \omega_{k-j+l} t \,+\, 
A \,\sin \omega_{j+l} t \,\sin \omega_{k-j+l} t 
\, \Biggr] 
\label{doublesum}
\end{equation}
where 
\[ A = \frac{\cos \omega_k \,-\, \cos \omega_{j+l} \,\cos \omega_{k-j+l}}
            {\sin \omega_{j+l} \,\sin \omega_{k-j+l}} 
\] 
The reader can check that this gives $\pt_t(0)=1$ for the trivial
Fourier component where $k=0$, and $\pt_t(n)=(-1)^t$ for the parity term
where $k=n$.

Using the identities $\cos a \,\cos b = (1/2)(\cos (a-b) + \cos
(a+b))$ and $\sin a \,\sin b = (1/2)(\cos (a-b) - \cos (a+b))$ we can
re-write Equation~\ref{doublesum} as
\begin{equation}
\pt_t(k) = \frac{1}{2^n} \sum_{j=0}^k \sum_{l=0}^{n-k}
\binom{k}{j} \binom{n-k}{l}
\left[
      \left( \! \frac{1-A}{2} \! \right) \cos \omega_+ t
    + \left( \! \frac{1+A}{2} \! \right) \cos \omega_- t
\right]
= \frac{1}{2^n} \sum_{j=0}^k \sum_{l=0}^{n-k}
\binom{k}{j} \binom{n-k}{l} \,Y
\label{doublesum2}
\end{equation}
where $\omega_\pm = \omega_{j+l} \pm \omega_{k-j+l}$.

The terms $\cos \omega_{\pm} t$ in $Y$ are rapidly oscillating with a
frequency that increases with $t$.  Thus, unlike the walk on the
cycle, the phase is rapidly oscillating everywhere, as a function of
either $l$ or $j$.  This will make the dominant contribution to
$\pt_t(k)$ exponentially small when $t/n=\pi/4$, giving us a small
variation distance when we sum over all $\vk$.

To give some intuition for the remainder of the proof, we pause here
to note that if Equation~\ref{doublesum2} were an integral rather than
a sum, we could immediately approximate the rate of oscillation of $Y$
to first order at the peaks of the binomials, where $j=k/2$ and
$l=(n-k)/2$.  One can check that $\domega_k/\dk \geq 2/n$ and hence
$\domega_+ / \dl = \domega_- / {\rm d}j \geq 4/n$.  Since $|A| \leq
1$, we would then write
\[ \pt_t(k) =_{\ord}
\frac{1}{2^n} \sum_{j=0}^k \sum_{l=0}^{n-k}
\binom{k}{j} \binom{n-k}{l}
\left( \e^{4ijt/n} \,+\, \e^{4ilt/n} \right)
\]
which, using the binomial theorem, would give
\begin{equation}
\left| \pt_t(k) \right| =_{\ord}
\left| \frac{1+\e^{4it/n}}{2} \right|^k \,+\,
\left| \frac{1+\e^{4it/n}}{2} \right|^{n-k}
= \cos^k \frac{2t}{n} \,+\,
  \cos^{n-k} \frac{2t}{n} 
\label{wouldgive}
\end{equation}
In this case the Diaconis-Shahshahani bound and the binomial theorem
give
\[
\norm{P_t - U}^2 \leq \frac{1}{4} \sum_{0 < k < n} \binom{n}{k} 
  \left( \cos^k \frac{2t}{n} \,+\, \cos^{n-k} \frac{2t}{n} \right)^2
\leq \frac{1}{2} \left[ \left(2 \cos^2 \frac{2t}{n}\right)^n +
  \left(1 + \cos^2 \frac{2t}{n}\right)^n - 1\right]
\]
If we could take $t$ to be the non-integer value $(\pi/4)n$, these
cosines would be zero.

This will, in fact, turn out to be the right answer.  But since
Equation~\ref{doublesum2} is a sum, not an integral, we have to be
wary of \emph{resonances} where the oscillations are such that the
phase changes by a multiple of $2\pi$ between adjacent terms, in
which case these terms will interfere constructively rather than
destructively.  Thus to show that the first-order oscillation indeed
dominates, we have a significant amount of work left to do. The
details of managing these resonances can be found in
Appendix~\ref{app:meat}.  The process can be summarized as follows:
\textsl{i.)} we compute the Fourier transform of the quantity $Y$ in
Equation~\ref{doublesum2}, since the sum of Equation~\ref{doublesum2}
can be calculated for a single Fourier basis function using the
binomial theorem; \textsl{ii.)} the Fourier transform of $Y$ can be
asymptotically bounded by the method of stationary phase. The dominant
stationary point corresponds to the first-order oscillation, but there
are an infinite number of other stationary points as well; so
\textsl{iii.)} we use an entropy bound to show that the contribution
of the other stationary points is exponentially small.

\begin{figure}
  \centering
  \mbox{\subfigure[Variation distance at time $t$ as a function of $t/n$.]{\label{subfigure:vardist}%
      \psfig{figure=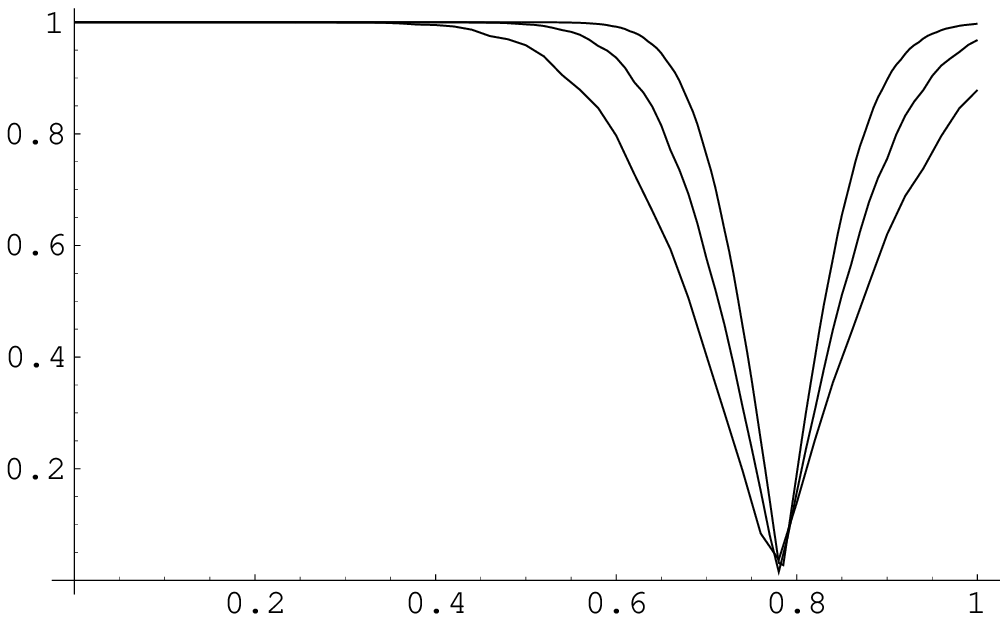,width=3in}}\quad%
    \subfigure[Probability as a function of Hamming weight.]{\label{subfigure:plateau}%
      \psfig{figure=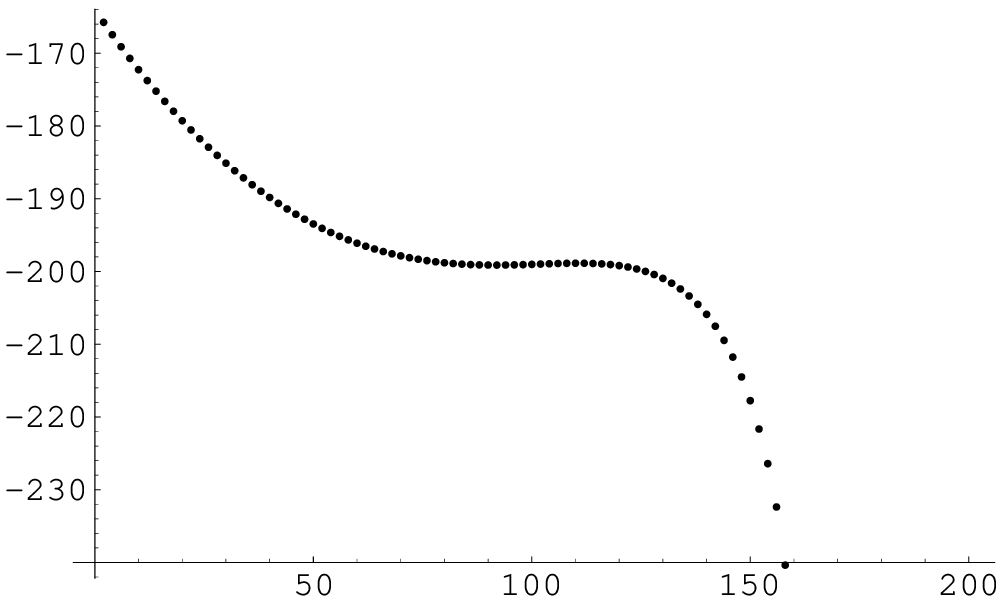,width=3in}}}

  \caption{\small Graph~(a) plots an exact calculation of the total
    variation distance after $t$ steps of the quantum walk for
    hypercubes of dimension 50, 100, and 200, as a function of $t/n$.
    At $t/n = \pi/4$ the variation distance is small even though the
    walk has not had time to cross the entire graph. This happens
    because the distribution is roughly uniform across the equator of
    the $n$-cube where the vast majority of the points are located.
    Graph~(b) shows the probability distribution on the
    $200$-dimensional hypercube after $157 \approx (\pi/4)n$
    steps. The probability distribution has a plateau of $2^{-199}$ at
    the equator, matching the uniform distribution up to parity.
    Shown is the log of the probability as a function of Hamming
    distance from the starting point.}  \label{figure:graphs}
    \vspace{.5mm} \hrule \end{figure}

To illustrate our result, we have calculated the probability
distribution, and the total variation distance from the uniform
distribution (up to parity), as a function of time for hypercubes of
dimension 50, 100, and 200.  In order to do this exactly, we use the
walk's permutation symmetry to collapse its dynamics to a function
only of Hamming distance.  In Figure~\ref{subfigure:vardist} we see
that the total variation distance becomes small when $t/n=\pi/4$, and
in Figure~\ref{subfigure:plateau} we see how the probability
distribution is close to uniform on a ``plateau'' across the
hypercube's equator.  Since this is where the vast majority of the
points are located, the total variation distance is small even though
the walk has not yet had time to cross the entire graph.

\noindent

\section{The continuous-time walk}
\label{sec:continuous-walk}

In this section we prove Theorem~\ref{thm:continuous}.  Childs, Farhi
and Gutmann \cite{quant-ph::ChildsFG2001} define quantum walks in a
different way, in which the unitary operator is generated from a
Hamiltonian $H$ using Schr{\"o}dinger's equation.  If $H$ is simply
the adjacency matrix of the graph, then $U_t = \e^{iHt} = 1 + iHt +
(iHt)^2/2 + \cdots $ giving a walk in continuous time.  The amplitude
of making $s$ steps is the coefficient $(it)^s/s!$ of $H^s$, which up
to normalization is Poisson-distributed with mean $t$.  They point out
that this avoids the need to extend the Hilbert space of the particle
with a direction space, and to define some local operation on it such
as Grover's operator, in order to make the walk unitary.  While this
approach is less familiar in computer science, a quantum computer
which is allowed to evolve in continuous time according to a certain
Hamiltonian seems just as physically reasonable as one which uses a
clock to evolve in discrete time as traditional computers do.

In the case of the hypercube, this walk turns out to be particularly
easy to analyze.  The adjacency matrix, normalized by the degree, is
\begin{equation}
H(\vx,\vy) = \left\{ \begin{array}{ll}
1/n & d(\vx,\vy) = 1 \\
0   & d(\vx,\vy) \neq 1
\end{array} \right. 
\label{heq}
\end{equation}
where $d$ is the Hamming distance.  The eigenvectors of $H$ and $U_t$
are simply the Fourier basis functions: if $v_\vk(\vx) = (-1)^{\vk
\cdot \vx}$ then $H v_\vk = \left(1-2k/n\right) v_\vk$ and $U_t
\,v_\vk = \e^{it\left(1-2k/n\right)} \,v_\vk$ where we again use $k$
to denote the Hamming weight of $\vk$.  If our initial wave vector has
a particle at $\vx=(0,\ldots,0)$, then its initial Fourier spectrum is
uniform, and at time $t$ we have
\[ \pst_t(\vk) = 2^{-n/2} \,\e^{it\left(1-\frac{2k}{n}\right)}. \]
Again writing the probability $P$ as the convolution of $\ps$ with
$\ps^*$ in Fourier space, we have
\[ \pt_t(\vk) = \sum_{\vk'} \pst_t(\vk') \,\pst_t^*(\vk \oplus \vk') 
    = \frac{1}{2^n} \sum_{\vk'} 
      \,\e^{2it \left(|\vk \oplus \vk'| - k' \right)/n}  
\]
We write this as a sum over all possible overlaps $j$ between $\vk'$
and $\vk$, and overlaps $l$ between $\vk'$ and $\vk \oplus \vk'$.
Noting that $k'=j+l$ and $|\vk \oplus \vk'|=k-j+l$, this gives
\begin{equation}
\pt_t(k) = \frac{1}{2^n} \sum_{j=0}^k \sum_{l=0}^{n-k}
              \,\e^{2it(k-2j)/n} 
= \cos^k \frac{2t}{n} 
\label{contpt}
\end{equation}

Finally, the Diaconis-Shahshahani bound on the total variation
distance between $P_t$ and the uniform distribution is
\[ \norm{P_t-U}^2
\leq \frac{1}{4} \sum_{k=1}^n \binom{n}{k} \abs{\pt_t(k)}^2 
= \left( 1 + \cos^2 \frac{2t}{n}\right )^n - 1
\] 
Astonishingly, at $t=(\pi/4)n$ and its odd multiples, this gives a
total variation distance which is exactly zero, showing that if we
sample at these times the probability distribution is {\em exactly}
uniform.  Note that this is possible even when $t < n$ since the
continuous-time walk has some probability for taking more than $t$
steps (and, in fact, paths with different numbers of steps interfere
with each other).  Thus the continuous-time walk has the same mixing
time as the discrete-time one, but with such a beautiful conspiracy of
interference that every position has an identical probability. This
concludes the proof of Theorem~\ref{thm:continuous}. For an
alternative derivation based on hypercube's structure as a product
graph, see Appendix~\ref{app:continuous}.

\section{Average mixing times}
\label{sec:mixingperiods}

In this section we discuss the mixing time as defined in
\cite{stoc::AharonovAKV2001}, where we choose to stop the quantum walk
at a time $t$ uniformly distributed in the interval $[0,T]$.  As
mentioned in the Introduction, this gives a probability distribution
$\oP_T = (1/T) \sum_{t=0}^{T-1} P_t$.  Since the Fourier transform is
a linear operation, we can look at the Fourier transform of $\oP_T$
instead.  In the case of the symmetric discrete-time walk,
Equation~\ref{doublesum2} shows that for $k > 0$, the Fourier
coefficient of $\oP_T$ consists of a sum of oscillating terms
proportional to $\cos \omega_\pm t$.  As $T \to \infty$, these
oscillations cancel, so we are left with just the constant term $k=0$
and $\oP_T$ indeed approaches the uniform distribution.

One could calculate an average mixing time for the symmetric
discrete-time walk using the methods of Appendix~\ref{app:meat}.  We
do not do that here.  However, we will now show that the general bound
of~\cite{stoc::AharonovAKV2001} predicts an average mixing time for
the $n$-cube which is exponential in $n$. The authors of that paper
showed that the variation distance between $\oP_T$ and the uniform
distribution (or more generally, the limiting distribution $\lim_{T
\to \infty} \oP_T$) is bounded by a sum over distinct pairs of
eigenvalues,
\begin{equation}
\norm{\oP_T - U} \,\leq\, \frac{2}{T} 
    \sum_{i,j \;{\rm s.t.}\; \lambda_i \neq \lambda_j} 
    \frac{|a_i|^2}{\left| \lambda_i - \lambda_j \right|} 
\label{aakvbound}
\end{equation}
where $a_i = \langle \psi_0 | v_i \rangle$ is the component of the
initial state along the eigenvector $v_i$.  Since this bound includes
eigenvalues $\lambda_j$ for which $a_j = 0$, we note that it also
holds when we replace $|a_i|^2$ with $|a_i a_j^*|$, using the same
reasoning as in~\cite{stoc::AharonovAKV2001}.

For the quantum walk on the cycle of length $n$, this bound gives an
average mixing time of $\ord(n \log n)$.  For the $n$-cube, however,
there are exponentially many pairs of eigenvectors with distinct
eigenvalues, all of which have a non-zero component in the initial
state.  Specifically, for each Hamming weight $k$ there are
$\binom{n}{k}$ non-trivial eigenvectors each with eigenvalue $\e^{i
  \omega_k}$ and $\e^{-i \omega_k}$.  These complex conjugates are distinct
from each other for $0 < k < n$, and eigenvalues with distinct $k$ are
also distinct.  The number of distinct pairs is then
\[ \sum_{k=1}^{n-1} \binom{n}{k}^2
   \,+\, 4 \sum_{k,k'=0}^n \binom{n}{k} \binom{n}{k'}
   = \Omega(4^n)
\]
Taking $|a_k| = 2^{-n/2} / \sqrt{2}$ from Equation~\ref{ai} and the
fact that $|\lambda_i - \lambda_j| \leq 2$ since the $\lambda_i$ are
on the unit circle, we see that Equation~\ref{aakvbound} gives an
upper bound on the $\eps$-average mixing time of size
$\Omega(2^n/\eps)$.  In general, this bound will give a mixing time of
$\Omega(M/\eps)$ whenever the initial state is distributed roughly
equally over $M$ eigenvectors, and when these are roughly equally
distributed over $\omega(1)$ distinct eigenvalues.

For the continuous-time walk, on the other hand, Equation~\ref{contpt}
shows that $\oP_T$ approaches the average of $\cos^k 2t/n$. In fact,
it is equal to this average whenever $T$ is a multiple of $(\pi/2) n$.
For $k$ odd this average is zero, but for $k$ even it is
\[ \frac{1}{\pi} \int_0^\pi \dx \,\cos^k x
 = \frac{2^k \pi}
        {\Gamma\!\left(\frac{1}{2} - \frac{k}{2} \right)^2 \, k!}
\]
Since these Fourier coefficients do not vanish, $\oP_T$ does not
approach the uniform distribution even in the limit $T \to \infty$.
In particular, the Fourier coefficient of $\oP_T$ for $k=2$ is
\begin{equation}
  \label{eq:big-coefficient}
  \widetilde{\oP}_T(2) = \frac{1}{T} \int_0^T \dt \,\cos^2 \frac{2t}{n}
  = \frac{1}{2} + \frac{\sin \,4T/n}{8T/n}
\end{equation}
This integral is minimized when $T=1.12335 \,n$, at which point
$\widetilde{\oP}_T(2) = 0.39138+$.  Since $\widetilde{\oP}_T(2)$ is
bounded below by this, it is easy to show that the total variation
distance $\norm{\oP_T - U}$ is bounded away from zero as a result.  Thus
there exists $\eps > 0$ such that no $\eps$-average mixing time
exists.

{\bf Acknowledgments.}  We are grateful to Dorit Aharonov, Mark
Newman, Tony O'Connor, Leonard Schulman, and Umesh Vazirani for
helpful conversations, and to McGill University and the Bellairs
Research Institute for hosting a conference at which a significant
part of this work was done.  This work is partially supported by NSF
grant PHY-0071139.


\newcommand{\etalchar}[1]{$^{#1}$}

\appendix
\section{Grover's diffusion operator}
\label{app:grover}

In general, the selection of the local operator $D$ on the
direction space appears to introduce a certain amount of
artificiality into the definition of a discrete-time quantum walk.
If we ask, however, that the operator obey the permutation
symmetry of the hypercube, then there is a one-parameter family of
such unitary operators up to multiplication by an overall phase.

To see this, suppose $D$ is unitary and permutation-symmetric.  Then
it can have only two distinct entries, namely those on the diagonal
and off it.  Let $D_{ij} = a$ if $i=j$ and $b$ if $i \neq j$.  Then
unitarity requires that $|a|^2 + (n-1) |b|^2 = 1$ and $2 \,{\rm
Re}\,ab^*) + (n-2)|b|^2 = 0$.  The first of these two equations
describes a circle, and their difference gives another, $|a-b|^2 = 1$.
The intersection of these circles gives at most two values for $b$
which differ only by a phase (and by conjugation if $a$ is real).
Solutions exist when $1-2/n \leq |a| \leq 1$.

To show that Grover's operator is the member of this family farthest
from the family of diagonal unitary matrices $\{c \id : |c|=1 \}$,
recall that the {\em operator norm} of a matrix $A$ is $\norm{A} =
{\rm Tr}\,A^\dagger A$.  Then the distance from $D$ to this family is
\[ \norm{D-c\id} = n|a-c|^2 + (n^2-n) |b|^2 = 2n(1-{\rm Re}\,ac^*) \]
When $c$ has the same phase as $a$ this is minimized at $2n(1-a)$,
and this minimum is maximized when $|a| = 1-2/n$.  This
corresponds to Grover's operator times an overall phase; in this
paper we take $a$ to be real and negative.

\section{Resonances in the discrete-time walk}
\label{app:meat}

In order to evaluate Equation~\ref{doublesum2}, we use Fourier
analysis again --- this time on functions of $j$ and $l$, or rather on
the rescaled variables
\[ x = \cos \omega_j = 1 - \frac{2j}{n}, \quad
   y = \cos \omega_l = 1 - \frac{2l}{n} 
\] 
We Fourier transform the quantity $Y$ in Equation~\ref{doublesum2}.
Since we are interested in oscillations of frequency $\Theta(t)$, we
write
\begin{equation}
 Y(x,y) = \sum_{p_x,\,p_y \in \Z} \yt\left(\frac{\pi p_x}{t},\frac{\pi
          p_y}{t}\right) \,\e^{-i \pi (p_x x + p_y y)}
\end{equation}
so that as $t$ goes to infinity, we may treat this as the integral
\begin{equation}
  Y(x,y) = \iint \yt(\beta_x,\beta_y) \,\e^{-it(\beta_x x + \beta_y
  y)} \,\dbeta_x \,\dbeta_y.  \label{yfou}
\end{equation}
Then, using the binomial theorem, we have
\begin{equation}
   \pt_t(k) = \iint \,\dbeta_x \,\dbeta_y \,\yt(\beta_x,\beta_y) 
   \,\e^{-it \left(\left(1-\frac{k}{n}\right) \beta_x 
                + \frac{k}{n} \beta_y \right)} 
   \,\cos^k \frac{\beta_x t}{n} 
   \,\cos^{n-k} \frac{\beta_y t}{n}
\label{binomial}
\end{equation}
We will show that $\yt$ peaks at values of $\beta_x$ and $\beta_y$
corresponding to the first-order oscillation, namely $(\beta_x,
\beta_y) = (2,0)$ and $(0,2)$.  This gives a form similar to
Equation~\ref{wouldgive}, so that if $2t/n = \pi/2$ the total
variation distance will be exponentially small.

We calculate $\yt$ by inverting Equation~\ref{yfou},
\[ \yt(\beta_x, \beta_y) = \frac{1}{4} \int_{-1}^{+1} \int_{-1}^{+1}
   \,\dx \,\dy \,Y(x,y) \,\e^{it (\beta_x x + \beta_y y)}
\]
where the normalization is due to the range of $x$ and $y$.  We divide
this integral into two terms, both of which are of the form
\begin{equation}
\iint \,\dx \,\dy
\left( \! \frac{1 \mp A}{2} \! \right) \,\cos \omega_\pm t
\,\,\e^{it (\beta_x x + \beta_y y)} 
= \ord\left( \iint \,\dx \,\dy  
\left( \! \frac{1 \mp A}{2} \! \right)
\,\e^{it (\omega_\pm + \beta_x x + \beta_y y)} 
\right)
\label{int}
\end{equation}
We can evaluate the right-hand integral in Equation~\ref{int} using
the method of stationary phase, also known as steepest descent, which
Nayak and Vishwanath \cite{quant-ph::NayakV2000} use to find the
asymptotic form of the wave function on the line.  In general, if $f$
is a slowly varying function then the asymptotic integral
$$
\lim_{t \to \infty} \iint f(x,y) e^{it\phi(x,y)}\dx\dy
$$
is dominated by contributions from the points $(x,y)$ in the domain
of integration where $\phi$ has zero gradient. (See, e.g.,
\cite{BleisteinH:Asymptotic}.)  If $r$ is the smallest integer such
that the $r$th derivative of $\phi$ at $(x,y)$ is nonzero, we say that
$(x,y)$ is {\em $r$th-order}.  In general, such asymptotic integrals
are dominated by contributions from the stationary points of highest
order.

In Equation~\ref{int} the slowly varying function is $(1 \mp A)/2$,
and the phase function is
\[ \phi_\pm(x,y) = \omega_\pm + \beta_x x + \beta_y y \]
Its derivatives are
\begin{eqnarray*}
\frac{\p \phi_\pm}{\p x} & = & - \frac{1}{\sin \omega_{j+l}}
                      \pm \frac{1}{\sin \omega_{k-j+l}} + \beta_x \\
\frac{\p \phi_\pm}{\p y} & = & - \frac{1}{\sin \omega_{j+l}}
                      \mp \frac{1}{\sin \omega_{k-j+l}} + \beta_y
\end{eqnarray*}
For both $\phi_+$ and $\phi_-$, setting these to zero gives four
stationary points $(x_0,y_0)$, where the angles $\omega_{j+l}$,
$\omega_{k-j+l}$ are described by
\begin{eqnarray}
\label{sin-cos}
\sin \omega_{j+l} = \frac{2}{\beta_x+\beta_y}
& & 
\sin \omega_{k-j+l} = \frac{2}{\left| \beta_x-\beta_y \right|} 
\label{cosines} \\
\cos \omega_{j+l} = x_0 + y_0 - 1 
= \pm \sqrt{1 - \left(\frac{2}{\beta_x+\beta_y}\right)^2}
& &
\cos \omega_{k-j+l} = 1 - \frac{2k}{n} - x_0 + y_0 
= \pm \sqrt{1 - \left(\frac{2}{\beta_x-\beta_y}\right)^2}
\nonumber
\end{eqnarray}
Note that the signs of the cosines can be chosen independently, and
all four possibilities exist for both $\phi_+$ and $\phi_-$.  Choosing
both cosines to be positive gives
\begin{eqnarray}
x_0 & = & \frac{1}{2} \left(
  \sqrt{\,1-\left(\frac{2}{\beta_x+\beta_y}\right)^2}
- \sqrt{\,1-\left(\frac{2}{\beta_x-\beta_y}\right)^2} \right) 
+ 1 - \frac{k}{n} 
\nonumber \\
y_0 & = & \frac{1}{2} \left(
  \sqrt{\,1-\left(\frac{2}{\beta_x+\beta_y}\right)^2}
+ \sqrt{\,1-\left(\frac{2}{\beta_x-\beta_y}\right)^2} \right)
+ \frac{k}{n}
\label{stationarypoints} 
\end{eqnarray}
The other three solutions are given by choosing one or both of the
cosines in Equation~\ref{cosines} to be negative, which affects the
signs of the square roots in Equation~\ref{stationarypoints}.  For
these solutions to be real, we require $\beta_y \geq |\beta_x|+2$ for
the stationary points of $\phi_+$, and $\beta_x \geq |\beta_y|+2$ for
the stationary points of $\phi_-$.  Thus $\beta_y-\beta_x \geq 2$ for
$\phi_+$ and $\beta_x-\beta_y \geq 2$ for $\phi_-$, and in both cases
$\beta_x+\beta_y \geq 2$.

To find the order of these stationary points, we calculate $\phi$'s second
derivatives at $(x_0,y_0)$:
\begin{eqnarray}
\frac{\p^2 \phi_\pm}{\p x^2} 
\;\;=\;\; \frac{\p^2 \phi_\pm}{\p y^2}
& = & - \frac{\cos \omega_{j+l}}{\sin^3 \omega_{j+l}}
    \mp \frac{\cos \omega_{k-j+l}}{\sin^3 \omega_{k-j+l}} 
\nonumber \\
\frac{\p^2 \phi_\pm}{\p x \,\p y} 
\;\;=\;\; \frac{\p^2 \phi_\pm}{\p y \,\p x}
& = & - \frac{\cos \omega_{j+l}}{\sin^3 \omega_{j+l}}
    \pm \frac{\cos \omega_{k-j+l}}{\sin^3 \omega_{k-j+l}}
\label{secondderivatives}
\end{eqnarray}
Given the restrictions on $\beta_x$ and $\beta_y$ for the stationary
points to be real, for each of $\phi_+$ and $\phi_-$ the second
derivatives are zero at exactly one pair of frequencies, namely
$\beta_x = 0$ and $\beta_y = 2$ for $\phi_+$, and $\beta_x = 2$ and
$\beta_y = 0$ for $\phi_-$.  We will call these the {\em dominant
stationary points}.  Note that at these frequencies we have
$\omega_{j+l} = \omega_{k-j+l} = \pi/2$ and the four stationary points
coincide at the peak of the binomials in Equation~\ref{doublesum2}
where $j=k/2$ and $l=(n-k)/2$.  Moreover, these frequencies are
exactly the first-order oscillations of $Y$ appearing in
Equation~\ref{wouldgive}.

Computing the third order derivatives at $\omega_{j+l} =
\omega_{k-j+l} = \pi/2$ gives
\begin{eqnarray*}
\frac{\p^3 \phi_{\pm}}{\p x^3} \;\;=\;\; 
\frac{\p^3 \phi_{\pm}}{\p x \,\p y^2}  
& = & - \left[ \frac{1}{\sin^3 \omega_{j+l}} 
             + \frac{3 \cos^2 \omega_{j+l}}{\sin^5 \omega_{j+l}} \right]
    \pm \left[ \frac{1}{\sin^3 \omega_{k-j+l}}
             + \frac{3 \cos^2 \omega_{k-j+l}}{\sin^5 \omega_{k-j+l}} \right]
\;\;=\;\; -1 \pm 1 \\
\frac{\p^3 \phi_{\pm}}{\p y^3} \;\;=\;\; 
\frac{\p^3 \phi_{\pm}}{\p x^2 \,\p y} 
& = & - \left[ \frac{1}{\sin^3 \omega_{j+l}} 
             + \frac{3 \cos^2 \omega_{j+l}}{\sin^5 \omega_{j+l}} \right]
    \mp \left[ \frac{1}{\sin^3 \omega_{k-j+l}}
             + \frac{3 \cos^2 \omega_{k-j+l}}{\sin^5 \omega_{k-j+l}} \right]
\;\;=\;\; -1 \mp 1 
\end{eqnarray*}
Thus the dominant stationary points are third order, and in their
vicinity $\phi_\pm$ takes the form
\[ \phi_\pm = \frac{1}{6} \left( -(x+y)^3 \pm (x-y)^3 \right)
   + \ord(x^4, y^4) 
\] 
Thus if we rotate $\pi/4$ to new variables $a = x+y$ and $b = x-y$, we
transform $\phi$ into the sum of two decoupled functions in the
vicinity of the dominant stationary point, and write the integral of
Equation~\ref{int} as the product of two one-dimensional integrals.
For one-dimensional integrals with a third-order stationary point
$x_0$, this takes the form~\cite[{\S}7]{BleisteinH:Asymptotic}
\[
\lim_{t \to \infty} \int \dx \,f(x) \,\e^{i t \phi(x)}  
= \frac{\Gamma(1/3)}{t^{1/3}} \,f(x_0) \,\e^{i t \phi(x_0)} 
  \,\frac{\e^{i \pi \,\sgn(\sigma)/6}}{3 \abs{\sigma}^{1/3}}
  \,+\, o(t^{-1/3})
\]
where $\sigma = \phi'''(x_0)$ is the third derivative at $x_0$.  Since
we have the product of two such integrals, and since $f(x_0) = (1 \mp
A)/2 = \ord(1)$ and $|\sigma| = 2$, the contribution of the dominant
stationary point to $\pt_t(k)$ is
\begin{equation}
 \left[\pt_t(k)\right]_{\rm dominant}
 = \ord\left(t^{-2/3} \left(\cos^k \frac{2t}{n} + \cos^{n-k} \frac{2t}{n}
       \right) \right)
\label{ptdominant}
\end{equation}

We now need to calculate the contribution of the other stationary points.
These are second-order, and their contribution takes the form
\begin{equation} 
\lim_{t \to \infty} \iint \,\dx \,\dy \,f(x,y) \,\e^{it \phi(x,y)} 
   = \frac{2\pi}{t} \sum_{(x,y)} f(x,y) \,\e^{it \phi(x,y)}
     \frac{\e^{i \pi \delta_{x,y} / 2}}
          {\sqrt{\left|\det \,\p^2 \phi_{x,y}\right|}} 
   + \ord\left(\frac{1}{t^2}\right)
\label{secondorder}
\end{equation}
where $\p^2 \phi_{x,y}$ is the matrix of second derivatives of $\phi$
at $(x,y)$, and $\delta_{x,y}$ is $+1$, $0$, or $-1$ depending on
whether zero, one, or both of its eigenvalues are negative.  From
Equation~\ref{secondderivatives} we have
\[ \det \,\p^2 \phi_\pm 
= \pm 4 \,\frac{\cos \omega_{j+l} \,\cos \omega_{k-j+l}}  
               {\sin^3 \omega_{j+l} \,\sin^3 \omega_{k-j+l}}
\]
Focusing on the oscillating part of Equation~\ref{binomial}, we have
\begin{equation}
   \iint \,\dbeta_x \,\dbeta_y
   \,\e^{i t \psi_\pm(\beta_x, \beta_y)} 
   \,\cos^k \frac{\beta_x t}{n} 
   \,\cos^{n-k} \frac{\beta_y t}{n}
\label{binomial2}
\end{equation}
where
\[ \psi_\pm(\beta_x, \beta_y) = \phi_\pm(x_0,y_0) 
   \,-\, \left(1-\frac{k}{n}\right) \beta_x
   \,-\, \frac{k}{n} \beta_y
\]
Since this really is an integral in the limit $n \to \infty$, the
$\cos^k$, $\cos^{n-k}$ terms create sharper and sharper peaks where
$\beta_x$, $\beta_y$ are multiples of 4.  We can approximate $\psi$ at
each peak to first order as a function of $\beta_x$ and $\beta_y$.
For the stationary point of $\phi_\pm$ where the sign of both cosines is
positive, $\psi_\pm$ is given by
\[ \psi_\pm(\beta_x, \beta_y) 
    = \sin^{-1} \frac{2}{\beta_x+\beta_y}
\,-\, \sin^{-1} \frac{2}{\beta_x-\beta_y}
\,+\, \sqrt{\left(\frac{\beta_x+\beta_y}{2}\right)^2 - 1}
\,\pm\, \sqrt{\left(\frac{\beta_x-\beta_y}{2}\right)^2 - 1}
\]
Its derivatives with respect to $\beta_x$ and $\beta_y$ are
\begin{eqnarray}
\frac{\p\psi_\pm}{\p\beta_x} 
& = & \frac{1}{2} \left(
  \sqrt{1-\left(\frac{2}{\beta_x+\beta_y}\right)^2}
- \sqrt{1-\left(\frac{2}{\beta_x-\beta_y}\right)^2}
\right) 
= x_0 - \left(1-\frac{k}{n}\right) = \frac{k-2j_0}{n} 
\nonumber \\
\frac{\p\psi_\pm}{\p\beta_y}
& = & \frac{1}{2} \left(
  \sqrt{1-\left(\frac{2}{\beta_x+\beta_y}\right)^2}
+ \sqrt{1-\left(\frac{2}{\beta_x-\beta_y}\right)^2}
\right)
= y_0 - \frac{k}{n} = \frac{n-k-2l_0}{n}
\label{psiprime}
\end{eqnarray}
and similarly for the other stationary points $(x_0, y_0)$; we can also
derive this directly from the definition of $\psi_\pm$ and the fact
that we are at a stationary point of $\phi_\pm$.  In other words, the
derivatives of $\psi$ are proportional to the distance of the stationary
points off the binomial peaks.

The entire $(\beta_x,\beta_y)$-plane can be tiled with $4 \times 4$
squares centered on these peaks.  Integrating Equation~\ref{binomial2}
on one such tile, say around the peak $\beta_x = 4p$, $\beta_y = 4q$,
gives
\begin{eqnarray}
& & \int_{4p-2}^{4p+2} \int_{4q-2}^{4q+2}
   \,\dbeta_x \,\dbeta_y
   \,\e^{i t \left( \frac{\p\psi}{\p\beta_x} \beta_x
                  + \frac{\p\psi}{\p\beta_y} \beta_y \right)} 
   \,\cos^k \frac{\beta_x t}{n} 
   \,\cos^{n-k} \frac{\beta_y t}{n}
\nonumber \\
& = & \frac{\pi^2 n^2}{2^n t^2} \,
\binom{k}{\frac{1}{2}(k-n\frac{\p\psi}{\p\beta_x})}
\binom{n-k}{\frac{1}{2}(n-k-n\frac{\p\psi}{\p\beta_y})}
= \frac{\pi^2 n^2}{2^n t^2} \,
\binom{k}{j_0}
\binom{n-k}{l_0}
\nonumber \\
& =_\ord & \exp \left[\,n \left(
    \frac{k}{n} \,h\!\left(\frac{j_0}{k}\right)
    + \left(1-\frac{k}{n}\right) \,h\!\left(\frac{l_0}{n-k}\right)
    - \ln 2 
  \right) \right] = \exp(nZ)
\label{binomial3}
\end{eqnarray}
where $h(z) = -z \ln z - (1-z) \ln (1-z)$ is the entropy function.
Note that if the quantity $Z$ in Equation~\ref{binomial3} is less than
$-\ln \sqrt{2}$ for all stationary points other than the dominant
ones, then their contribution to $|\pt(k)|^2$ will be $2^{-\gamma n}$
where $\gamma > 1$, in which case summing over all $k$ will give an
exponentially small contribution, $\ord(2^{(1-\gamma)n})$, to the
total variation distance.  To confirm this, note that $Z$ is maximized
by the other stationary points closest to the origin, such as the
stationary point of $\phi_+$, with both cosines positive, where
$\beta_x=0$ and $\beta_y=4$.  From Equation~\ref{psiprime} this gives
$\p\psi/\p\beta_x = 0$ and $\p\psi/\p\beta_y = \sqrt{3}/2$, and so
$j_0 = k/2$ and $l_0=((1-\frac{\sqrt{3}}{2})n-k)/2$.  Both binomials
are non-zero only in the interval $k \in
\left(0,(1-\frac{\sqrt{3}}{2})n\right)$ and $Z$ is maximized at $k=0$,
where
\[ Z = h\left(\frac{1}{2}-\frac{\sqrt{3}}{4}\right) - \ln 2 = -0.447
     < \ln \frac{1}{\sqrt{2}} = -0.346 
\] 
The other second-order stationary points are this far or farther from
the origin, giving values of $j_0$ and $l_0$ farther off the binomial
peaks, and therefore smaller entropies.  

Recalling Equation~\ref{secondorder} above, our final concern is the
sum of the heights of these peaks,
\[
\sum_{\beta_x, \beta_y} 
  \frac{1}{\sqrt{\abs{\det \p^2 \phi_{\beta_x,\beta_y}}}}
\]
taken over all second-order stationary points $(\beta_x,\beta_y)$.
Since these occur when $\beta_x, \beta_y$ are multiples of 4, from
Equation~\ref{sin-cos} we have $\abs{\cos \omega_{j+l} \cos
\omega_{k-j+l}} \geq 3/4$.  Then
\[ \abs{\det \p^2 \phi_{\pm}(\beta_x,\beta_y)} 
\geq \frac{3}{\abs{\sin^3 \omega_{j+l} \sin^3 \omega_{k-j+l}}}
= \frac{3}{64} \,\abs{\beta_x + \beta_y}^3 
               \,\abs{\beta_x - \beta_y}^3
\]
and it is sufficient to show that the sum
\[ \sum_{\beta_x \neq \beta_y} 
   \,|\beta_x + \beta_y|^{-3/2} 
   \,|\beta_x - \beta_y|^{-3/2}
\]
converges.  Again rotating by $\pi/4$ to variables $a = \beta_x +
\beta_y$ and $b = \beta_x - \beta_y$, we get the sum
\[ \sum_{a,b} |a|^{-3/2} |b|^{-3/2} 
\leq \left( \sum_a |a|^{-3/2} \right)^2 
\] 
Observing that $\sum_{a > 0} a^{-3/2}$ converges shows that the
contribution of the second-order stationary points is exponentially
small.

Now we return to the dominant contribution to $\pt_t(k)$,
Equation~\ref{ptdominant}.  If we could have $t = (\pi/4)n$ exactly,
this dominant term would be zero, leaving us with the second-order
stationary points and an exponentially small total variation distance.
However, in the discrete-time walk $t$ must be an integer.  Setting $t
= \lceil (\pi/4) n \rceil$, we have $\cos 2t/n = \ord(1/n)$.  Using
the binomial theorem and Equation~\ref{ptdominant}, the
Diaconis-Shahshahani bound gives 
\begin{eqnarray*}
 \norm{P_t-U}^2 & =_\ord  & n^{-4/3} \sum_{0 < k < n} \binom{n}{k} 
    \left( 2 \cos^{2k} \frac{2t}{n} + 2 \cos^n \frac{2t}{n} \right) \\
& \leq & 2 n^{-4/3} \left[ \left( 2 \cos \frac{2t}{n} \right)^n
              + \left( 1 + \cos^2 \frac{2t}{n} \right)^n - 1 \right] 
\;=\; \ord(n^{-7/3})
\end{eqnarray*}
and so the total variation distance is $\norm{P_t-U} =
\ord(n^{-7/6})$, completing the proof of Theorem 1.

\section{A graph product derivation of the continuous-time walk}
\label{app:continuous}

As an alternate derivation for the continuous-time walk, we can
calculate the wave function $\psi_t$ directly by exploiting the
hypercube's simple structure as a product graph.  Let $\sigma_x$ be
the Pauli matrix ${\scriptsize \left( \!\begin{array}{cc} 0\! & \!1 \\ 
      1\!  & \!0 \end{array} \! \right)}$.  Then we can rewrite
Equation~\ref{heq} as
\[ H = \frac{1}{n} \sum_{j=1}^n \id \otimes \cdots \otimes \sigma_x
                        \otimes \cdots \otimes \id
\]
where the $j$th term in the sum has $\sigma_x$ appearing in the $j$th
place in the tensor product.  Then using the identity $(A \otimes B)(C
\otimes D) = AB \otimes CD$, and the fact that $\e^{A+B} = \e^A \e^B$
when $A$ and $B$ commute, we have
\[ U = \e^{iHt} = \prod_{j=1}^n \id \otimes \cdots \otimes 
               \e^{it \sigma_x / n} \otimes \cdots \otimes \id 
 = \left[ \e^{it \sigma_x / n} \right]^{\otimes n}
 = \left( \begin{array}{cc} \cos \,t/n & i \,\sin \,t/n \\
                            i \,\sin \,t/n & \cos \,t/n
   \end{array} \right)^{\otimes n}
\]
where $A^{\otimes n}$ is the tensor product of $n$ copies of $A$.  If
$\psi_0 = | 0 \cdots 0 \rangle = | 0 \rangle^{\otimes n}$, then
\[ \psi_t = U_t \psi_0 = \left[ 
   \left( \cos \frac{t}{n} \right) | 0 \rangle
 + \left( i \,\sin \frac{t}{n} \right) | 1 \rangle \right]^{\otimes n}
\]
and we see that the continuous-time walk is equivalent to $n$
non-interacting one-qubit systems.  Then the amplitude for observing
the particle at a position $\vx$ with Hamming weight $x$ is
\[ \psi_t(\vx) = \left( \cos \frac{t}{n} \right)^{n-x} 
                 \left( i \,\sin \frac{t}{n} \right)^x 
\]
which when $t = k(\pi/4)n$ for $k$ odd gives $|\psi_t(x)|^2 = 2^{-n}$,
the uniform distribution.

\end{document}